\documentclass[reprint,superscriptaddress,aps,showpacs,amsmath,amssymb]{revtex4-1}
\usepackage{graphicx}
\usepackage{dcolumn}
\usepackage{bm}
\usepackage{soul} 
\usepackage{cancel}
\usepackage{color}
\usepackage{color}
\definecolor{bluecolor}{rgb}{0,0,0.8}

\definecolor{redcolor}{rgb}{.7,0.,0.}

\input{epsf}

\begin{document}
\title{Dynamics of transposable elements generates \\ structure and symmetries in genetic sequences}

\vskip4mm                      

\author{Giampaolo Cristadoro}
\affiliation{Dipartimento di Matematica e Applicazioni, Universit\`a di Milano - Bicocca, Italy} 
\author{Mirko Degli Esposti}
\affiliation{Dipartimento di Informatica, Universit\`a di Bologna, Italy}
\author{Eduardo G. Altmann}
\affiliation{School of Mathematics and Statistics,
  University of Sydney, Australia}

\begin{abstract}
Genetic sequences are known to possess non-trivial composition together with symmetries in the frequencies of their components. 
Recently, it has been shown that symmetry and structure are hierarchically intertwined in DNA, suggesting a common origin for both features. However, the mechanism leading to this relationship is unknown. Here we investigate a biologically motivated dynamics for the evolution of genetic sequences. We show that a metastable (long-lived) regime emerges in which sequences have symmetry and structure interlaced in a way that matches that of extant genomes.
\end{abstract}

\maketitle

\paragraph{Introduction.}
Transposable elements (TEs) are DNA sequences that can relocate themselves in new sites of the genome.  They were firstly discovered in maize by  B.~McClintock in the mid-1940s and initially  considered as  parasites with no functional roles \cite{MC50}. Nowadays TEs are known to be ubiquitous in both prokaryotes and eukaryotes genomes \cite{FP07,K81} and little  doubts  are left of their prominent  role in genome evolution,  shaping  structure and function in a multitude of ways  \cite{BBGGetal18, F12}.  As TEs constitute more than half of the  sequence in many higher eukaryotes,  a fingerprint of their presence can be quantitatively extracted from  the statistical properties of their host DNA. Indeed,  TEs properties were shown to be crucial  in explaining structural global features of genome sequences \cite{HGBSH, SRMA16, MA13,MAL05,HJ96, BGHPSS93}.

Recently, Albrecht-Buehler~\cite{B06} suggested that TEs were the main driving force for the emergence of the second Chargaff parity rule. This rule states that, in {\it each} strand of the DNA, the frequencies of a short oligonucleotide $\mathbf{w}$ is approximately equal to that of its symmetrically related $\mathbf{\hat{w}}$, obtained from $\mathbf{w}$ by reversing the order of the symbols and substituting each nucleotide with its conjugated $A ~\leftrightarrow~T$  and $C~\leftrightarrow~G$ (e.g. $ \mathbf{w}=ACTGGCT$, $\hat{\mathbf{w}}=AGCCAGT$). It has been first observed by Chargaff in the 1950s~\cite{RKC68} and since then detected across different  organisms leading to  different proposals for its origin and function~\cite{R91, MB06,NA06,QC01,FTW92,P93,BF99,PHB02,KFCHZZL09,ABGRPF13,BF99b,LL99,ZH10,HMO12,CBHDK19,FTPM19}. The importance of Albrecht-Buehler explanation is that it shows how this symmetry  naturally emerges as an  {\it asymptotic} outcome of the cumulative action of inversions/transpositions, one of the main mechanism of relocation of TEs. As we will show,  while the proposed mechanism nicely  induces  Chargaff  symmetry  in the asymptotic DNA, it does it at the cost of trivialisation of the structural properties of the sequence:  symmetry is  obtained because of the complete randomization of the full double-stranded DNA. 
\begin{figure}[h!]
\begin{center}
\includegraphics[width=\columnwidth]{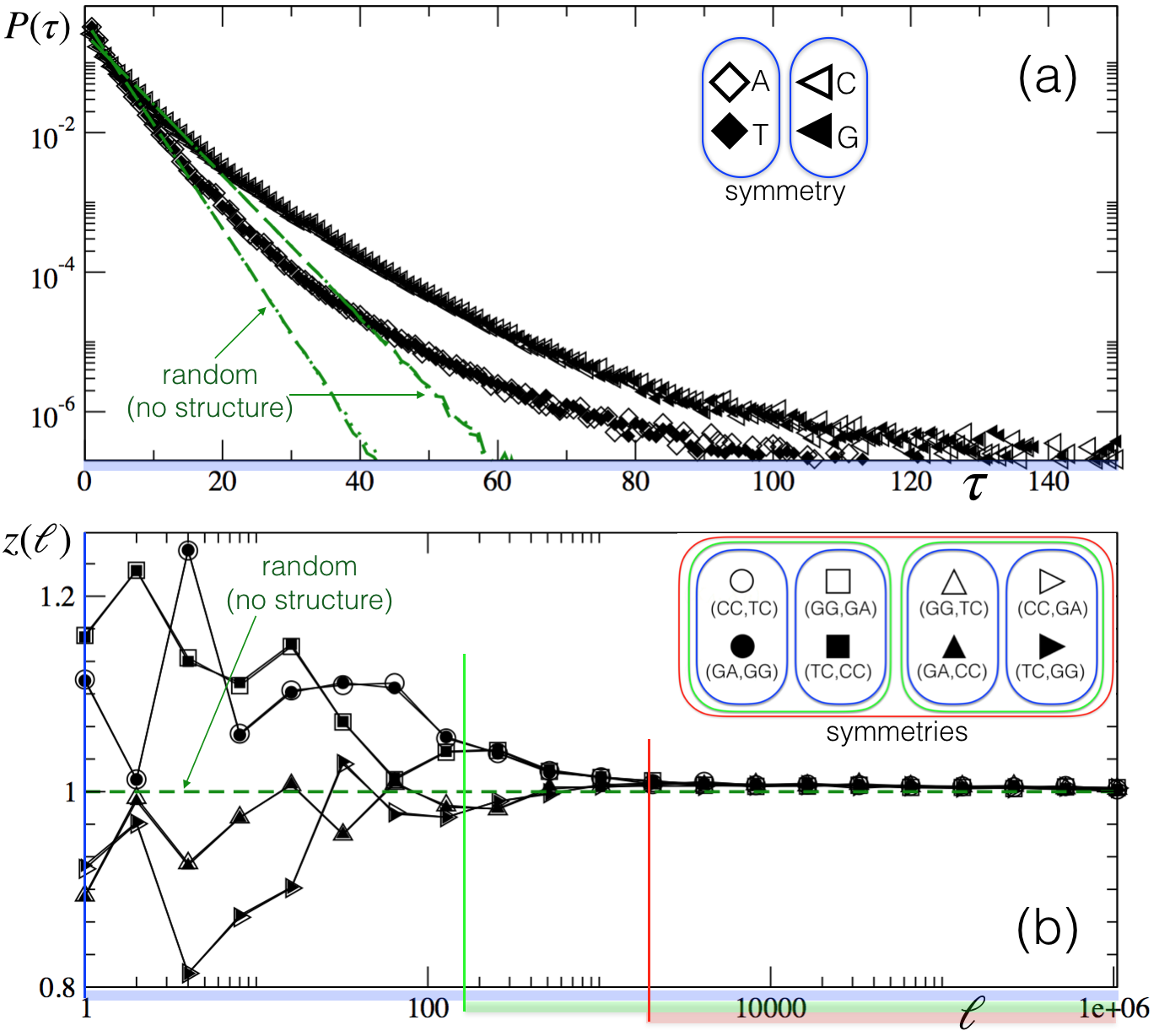}
\caption{
{\bf{Symmetry and structure are intertwined in DNA.}}
Results are shown for Homo-Sapiens chromosome 1 (symbols) and its randomly shuffled version (dashed lines). Each curve corresponds to one observable. Symmetrically related observables appear in the same box in the legend. 
\emph{(a)} Distribution $P(\tau)$ of recurrence-times $\tau$ (measured in number of basis) between successive occurrences of the same nucleotide.
 \emph{(b)} Probability $f_{X_A,X_B}(\ell)$ that the bigrams $X_A$ and $X_B$ appear separated by a distance $\ell$. Plotted is the normalized cross-correlation $z_{[X_A,X_B]}(\ell)=f_{X_A,X_B}/(f_{X_A}f_{X_B})$ as a function of $\ell$, for symmetrically related couples $[X_A,X_B]$(see legend). Different nested symmetries  are valid at different scales $\ell$ (see Ref.~\cite{CDEA18} for further details): for $\ell\lessapprox 150$  Chargaff  $z_{[X_A,X_B]}=z_{[\hat{X}_B,\hat{X}_A]}$, for $150 \lessapprox \ell \lessapprox 1500$ Chargaff and reverse symmetry $z_{[X_A,X_B]}=z_{[X_B,X_A]}$, and for $\ell \gtrapprox 1500$  complement $z_{[X_A,X_B]}=z_{[\hat{X}_A,\hat{X}_B]}$ and reverse symmetry. %
}
\label{fig.1}
\end{center}
\end{figure}
In view of the ubiquity of  complex structures in genomes~\cite{
PBGHSSS92,LK92,V92,A92,P92,LMK94,ATVAMA01,FS12,CPTC15}, this result raises the question whether symmetry can appear without a full  randomization of the sequence and in a way that is compatible with the existence of structure. The importance of this question is enhanced by our recent findings~\cite{CDEA18} that Chargaff symmetry extends beyond the frequencies of short oligonucleotides -- remaining valid on scales where non-trivial structure is present --  and that an hierarchy of other symmetries exists, nested at different structural scales.  This findings are confirmed in Fig.~\ref{fig.1}, which shows how commonly used indicators of structures, such as recurrence-time distribution (panel a) and correlation functions (panel b), coincide for symmetrically related observables at different scales.

 In this work we present a biologically motivated dynamical process that explains the observed relation between  symmetry and structure in DNA sequences.
In particular, we propose a model that mimics the action (inversion/transpositions) of TEs on DNA and we analytically describe  its dynamical behavior. Using indicators to quantify both symmetry and the presence of non-trivial structure in symbolic sequences, we show that the  co-occurrence of symmetry and structure is an emergent statistical property in sequences generated by such model, reproducing the same hierarchical relation detected in extant genomes.

\paragraph{Quantifying structure and symmetry.} We consider symbolic sequences  $\mathbf{s}=\{s_i\}_{i=1}^N$  of length $|\mathbf{s}|=N$ with  $s_i \in \mathcal{A}=\{A,C,G,T\}$. Given a subsequence $\mathbf{a}$ of $\mathbf{s}$ (a word) we denote its corresponding reverse-complemented word  as $\hat{\mathbf{a}}$, obtained from $\mathbf{a}$ by reversing the order of the symbols and substituting each nucleotide $\{A,C,G,T\}$ by its complementary one $A ~\leftrightarrow~T$  and $C~\leftrightarrow~G$. 
We call $f_x(\mathbf{s})$ the percentage of the nucleotide $x$ in the sequence $\mathbf{s}$. Finally, we denote by $CG(\mathbf{s}):= f_C(\mathbf{s})+f_G(\mathbf{s})$ (the so called  CG-content).
 In the following, it will be useful to  partition  the full set $\mathcal{A}^N$  into disjoint subsets of fixed CG-content  $\mathcal{B}^N(k):=\{\mathbf{s} \in \mathcal{A}^N | CG(\mathbf{s})=k/N \}$;  $\mathcal{A}^N= \cup_{k=0}^N \mathcal{B}^N(k)$.

We introduce the following simple indicators of the presence of Chargaff Symmetry and of non-trivial structure composition of a  given sequence $\mathbf{s}$.

 To quantify the  compliance of $\mathbf{s}$ with Chargaff symmetry, we average the normalized difference of the abundance between a nucleotide and its symmetric one (see \cite{PHB02} where a similar measure was firstly introduced)
\begin{equation}\label{IndicatorSymmetry}
I_{sym}(\mathbf{s})= \frac14 \sum_{x\in \mathcal{A}} \frac{|f_x(\mathbf{s}) - f_{\hat{x}}(\mathbf{s})|}{f_x(\mathbf{s}) + f_{\hat{x}}(\mathbf{s})}. 
\end{equation}
$I_{sym}=0$ indicates a fully Chargaff-symmetric sequence, $I_{sym}=1$ is obtained for a sequence for which Chargaff is perfectly violated ($f_A=f_C=0.5,f_T=f_G=0$), and $I_{sym}=0.08$ is obtained for a $2\%$ variation of equal frequencies (e.g., $f_A=f_C=0.23, f_T=f_G=0.27$).  For simplicity, we consider $I_{sym}>0.08$ to be a  violation of Chargaff symmetry.

To quantify the presence of  non-trivial structures in a given symbolic sequence $\mathbf{s}$ we first compute the distribution $P(\tau)$ of distances $\tau$ between two successive occurrence of the same nucleotide $x$. For random sequences,  $P(\tau)$ decays exponentially as $P(\tau) = {f_x}{(1-f_x)^{\tau-1}}$ and thus has average $1/f_x$ and standard deviation $\sqrt{1-f_x}/f_x$ (which is $\approx 1/f_x$ for small $f_x$). In contrast, the presence of a fat tail (standard deviation much larger than the mean)  is considered a signature of a complex organization. We thus quantify structure as the distance of $\mathbf{s}$ from random sequences by
\begin{equation}\label{IndicatorStructure}
 I_{str}(\mathbf{s}) =   \frac14\sum_{x\in \mathcal{A}}    \left( \frac{1}{\sqrt{1-f_x(\mathbf{s})}}\frac{\sigma_{\tau}(x)}{\mu_{\tau}(x)}   -1 \right), 
\end{equation}
where $\mu_\tau \equiv \langle \tau \rangle $ and $\sigma_\tau\equiv \sqrt{\langle \tau^2 \rangle - \langle \tau \rangle^2}$ are the mean and standard deviation of the measured $P(\tau)$, and $\sqrt{1-f_x}$  is the expected $\sigma_\tau/\mu_\tau$ 
for nucleotide $x$ in a random sequence. For random sequence we thus have $I_{str}(\mathbf{s}) =0 $, while departure from this value mark the presence of non-trivial structure. For simplicity, we consider $I_{str}>0.01$ to be a signature of structure.

\paragraph{Dynamics.} 
 We investigate symmetry and structure of sequences that evolve through the following dynamics, that maps one sequence  $\mathbf{s}(t) \in \mathcal{A}^N$ into another sequence $\mathbf{s}({t+1}) \in \mathcal{A}^N$ by mimicking the action of TEs \cite{B06}. The dynamics is defined composing two actions:

\begin{itemize}
\item[(i)] pick a random position $j$ of $\mathbf{s}$ and a random size $\ell \ge 0$, with $\langle \ell \rangle = L$ \footnote{More precisely, the pairs $(j,\ell)$ are drawn, independently from previous iterations, from a joint distribution $\rho(j,\ell)$ chosen such that  its marginal  $\phi(\ell)$  has   support  contained in $[0,N]$, finite average $L$,  and the  conditional distribution of positions $\psi(j|\ell )$  is uniform in $[1,N-\ell+1]$. We consider distributions $\phi(\ell)$  that guarantee ergodicity of the Markov Chain. We expect ergodicity to be generically valid; e.g., it suffices to have non-zero probability for the identity  transformation (i.e. $\phi(0) \neq 0$)  and for single-nucleotide complementing ($\psi(1) \neq0$)}. 
\item[(ii)] replace the subsequence  $\mathbf{b}\equiv \{ s_i\}_{i=j}^{j+\ell -1}$ of size $\ell$ starting at position $j$, by its reverse complement $\hat{\mathbf{b}}$. 
  \end{itemize}
 The couple $(j,\ell)$ parametrizes  the effect of an inversion/transposition, which we denote by $g_{(j,\ell)}: \mathcal{A}^N  \to  \mathcal{A}^N$. Its action  has interesting  properties: $g_{(j,\ell)}$ is an involution for every $(j,\ell)$ and  the total number  of $C$ and $G$ (or, equivalently, of $A$ and $T$) is invariant under $g$: $CG(\mathbf{s}_{t})=CG(\mathbf{s}_{0}) \quad \forall t$. This implies that the dynamics is  restricted to the invariant subspace of sequences with constant  CG-content $\mathcal{B}^N(CG(\mathbf{s}_{0}))$.
  
\paragraph{Asymptotic  equilibrium.}
The dynamics can be equivalently described as  an ergodic Markov chain over the space of sequences $\mathcal{B}^N(CG(\mathbf{s}_{0}))$. The fact that   $g_{(j,\ell)}$ is an involution forces the transition matrix to be  bi-stochastic and thus  in the asymptotic equilibrium  all sequences are equiprobable. This  means  that, for $t\to \infty$ and irrespective of the initial ancient DNA sequence, the evolution asymptotically leads to  sequences that can be equivalently considered generated by an independent and identically distributed (iid) process with $p(G)=p(C)=CG(\mathbf{s}_{0})/2$ and $p(A)=p(T)=(1-CG(\mathbf{s}_{0}))/2$.  Therefore, the expected value of our indicators of symmetry and structure  Eqs.~(\ref{IndicatorSymmetry}) and~(\ref{IndicatorStructure}) vanish asymptotically
$$ \lim_{t\to\infty}  I_{str}(\mathbf{s}(t)) = \lim_{t\to\infty} I_{sym}(\mathbf{s}(t)) = 0,$$
for any  initial sequence~${\bf s}(0)$\footnote{We can intuitively understand this result by noting that each action of  the transposon effectively creates two cuts in the sequence and moves them by a distance $L$ on average. Since cuts can happen at any location,  this process  eventually mixes complementary basis at different positions and breaks  any correlations originally present in $\mathbf{s}(0)$.}.
This shows analytically that the TE dynamics asymptotically leads to Chargaff symmetric sequences, in agreement with previous claims \cite{B06}. However,  this symmetric equilibrium is a (trivial) consequence of a full randomization. Therefore our results show also that the current explanations of the second Chargaff parity rule~\cite{B06} is not satisfactory as it is not compatible with any structure, which is known to remain significant at distances of several thousands of nucleotides \cite{
PBGHSSS92,LK92,V92,A92,P92,LMK94,ATVAMA01,FS12,CPTC15}  (see also Fig.\ref{fig.1}). Next we show that  the same TE dynamics is rich enough by showing that symmetric sequences with non-trivial structure are generated  pre-asymptotically as long-lived metastable states of TEs dynamics.

\paragraph{Symmetry and structure over time - three regimes.}
We now  investigate  symmetry and structure of the sequences $\mathbf{s}(t)$ by computing how our indicators $I_{sym}$ an $I_{str}$ depend on time $t$ (i.e., their values after $t$ applications of $g_{(j,\ell)}$).
We  show that Chargaff symmetry emerges much before equilibrium,  together with a complex domain-like structure. 

We first investigate  structural properties of sequences after a finite number $t$ of iterations. 
We define a \emph{domain} of $\mathbf{s}(t)$ as a subsequence of consecutive sites that have been involved in the same series of reverse/complement events.  We then distinguish between domains of type  $\Gamma$ and $\hat{\Gamma}$, depending on whether the number of transformations $g$ they were involved is even or odd, respectively. By definition, the starting sequence is composed  by a single domain of type  $\Gamma$. After one iteration it is split into three domains, two of type $\Gamma$ and one of type $\hat{\Gamma}$ of length $\ell_1$, corresponding to the subsequence involved in the first  reverse/complement event. 
We now compute the average sizes  $\langle \ell_{\Gamma} \rangle (t)$ and $\langle \ell_{\hat{\Gamma}} \rangle (t)$  of domains after $t$ iterations. Three regimes can be identified:

\noindent(i) For short times $t$, if $L \ll N$,  the probability that the first few iterates all involve different subsequence is very high\footnote{Quantitatively, if we drop  $t \le N/L$ points uniformly at random on an interval of length $N$, they will be separated by a distance at least  $L$ with probability $~(1-(t-1)L/N)^t$}. At each iterate, a subsequence of a domain of type $\Gamma$ of average size $\langle \ell \rangle=L$ is created, cutting  a  domain of type $\hat{\Gamma}$. Thus we have that in this regime:
\begin{equation}
  \langle \ell_{\hat{\Gamma}} \rangle (t)= L \; \text{ and } \; \langle \ell_{{\Gamma}} \rangle (t) = N/t.
\end{equation}
This regime lasts until iterates start overlapping, which happens when $N/t \approx L$ and average domain-sizes equalize  $\langle \ell_{\hat{\Gamma}} \rangle (t)=\langle \ell_{{\Gamma}} \rangle (t)=L$. This regime is thus  valid for   $0 < t  \lesssim  t_{metastable}=N/L$. 

\noindent (ii) For $ t \gtrsim t_{metastable}=N/L$  a typical  reverse/complement event will overlap with  more than one domain. In this case all the domains that lie  fully inside the subsequence involved in the reverse/complement event will change type (and position) without changing length;  the domains  at the border are instead split in two sub-domains of different type. The randomness of this process guarantees that  the already reached balance between the number and average length of  the  two domains  types $\Gamma$ and $\hat{\Gamma}$ is not broken while their common average length decreases in time as
\begin{equation}
  \langle \ell_{\hat{\Gamma}} \rangle (t)=\langle \ell_{{\Gamma}} \rangle (t)=N/t.
\end{equation}
This second regime ends  after a number of iterations $t \sim t_{equilibrium}= N$ when equilibrium is reached.

\noindent (iii) For $t> t_{equilibrium} = N$ the average lengths stabilize at the stationary value
\begin{equation}
\langle \ell_{\hat{\Gamma}} \rangle (t)=\langle \ell_{{\Gamma}} \rangle (t)=1, 
\end{equation}
 and the sequence can be thought as a realization of the asymptotic equilibrium discussed above.
\begin{figure}[t!]
\begin{center}
\includegraphics[width=\columnwidth]{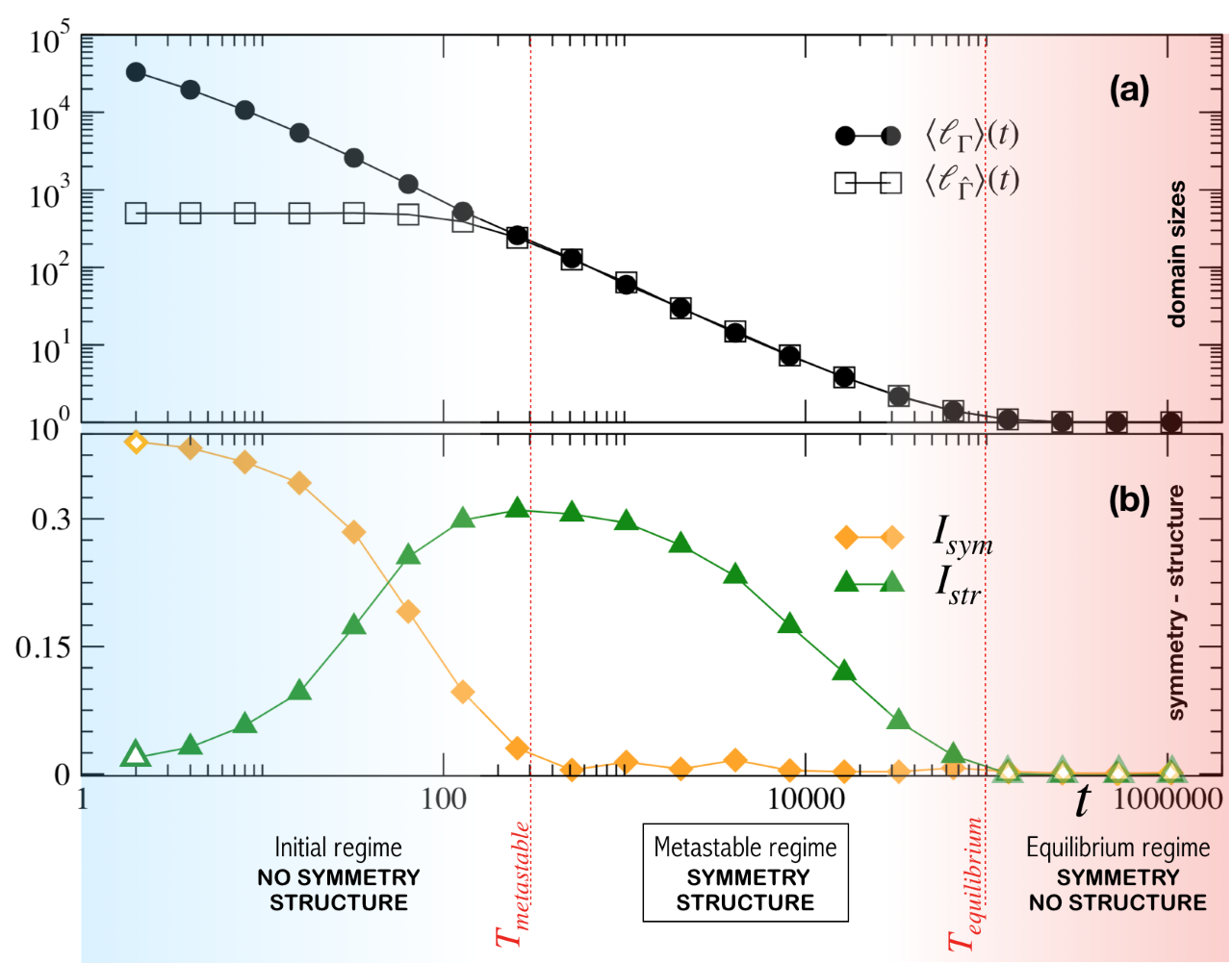}
\caption{{\bf{Temporal evolution of symmetry and structure in the model.}} \emph{(a)} Numerical evaluation of the average sizes of domains  of the two types $\langle \ell_{\Gamma}\rangle(t)$ and $ \langle \ell_{\hat{\Gamma}}\rangle(t)$ as a function of the number of iterates $t$ of TE's dynamics. \emph{ (b)} Numerical evaluation of the symmetry and structural properties of the sequence generated by the dynamics and quantified by the indicators  $I_{str}(t)$ and $I_{sym}(t)$.  The filled symbols in $I_{str}$ indicate that these values are statistically different from a random sequence ($p-value<0.01$, equivalent results are obtained using as an alternative definition of $I_{str}$ the Jensen-Shannon divergence between the $P(\tau)$ obtained in the model and in random sequences).
The sequences $\mathbf{s}(t)$ have length $N=10^5$ and the  size of reverse/complement events  is $L=500$, thus leading to time-scales $t_{metastable}=10^2$ and $t_{equilibrium}=10^5$. The starting sequence is  fully random with $f_A=0.1, f_C=0.2, f_G=0.3, f_T=0.4.$
}
\label{fig2}
\end{center}
\end{figure}

We now explain how structure $I_{str}(\mathbf{s}(t))$ and symmetry $I_{sym}({\bf s}(t))$ depend on the domain sizes $\langle \ell_\Gamma \rangle$ and $\langle \ell_{\hat{\Gamma}} \rangle$ and thus on the different regimes.

\begin{itemize}

  \item[$I_{str}(\mathbf{s})$]: in order to identify the contribution of the dynamics in generating complex structural features,  we consider an initial  $\mathbf{s}(0)$ generated by  an iid process (no structure, $I_{str}(0)=0$).  With this choice,  a value $I_{str} \neq 0$  signal  the construction, under the action of the dynamics, of different domain-types. In particular, at  $ t_{metastable}$  and for  $L >>1 $,  the total variance $\sigma_{\tau}^2$  can be estimated, using the law of total variance, as the sum of two components: one that measure variability of the mean of returns between domain-types and the other measuring variability  of returns within each type. Accordingly
   $I_{str}(t)$ grows from $0$ to the  value  $I_{str}(t_{metastable})>0$  at the end of the first regime. In the second regime the domain sizes decay and  $I_{str}(t)$ decreases to  zero at equilibrium (at $t_{equilibrium}$).  
 In terms of regimes we thus expect: (i) $I_{str}$ grows; (ii) $I_{str}$ decays; (iii) $I_{str}=0$.

\item[$I_{sym}({\bf s})$]: each domain of type $\Gamma$ is a subsequence of the ancient sequence $\mathbf{s}(0)$. If average size of such domains at time $t$  is large enough, the frequency of  each nucleotide are approximately the same as their  frequency in  $\mathbf{s}(0)$; similarly for $\hat{\Gamma}$ and  $\mathbf{\hat{s}}(0)$.
No  constraints are imposed  to the symmetry of the ancient genome. In particular,   if the original sequence is not Chargaff symmetric  $I_{sym}(\mathbf{s}(0))>0$ then the symmetry remains broken for all $t \lesssim t_{metastable}$ as quantified by
$I_{sym}(\mathbf{s}(t)) \simeq  \frac{t}N \left|  \langle \ell_{\Gamma}\rangle(t) - \langle \ell_{\hat{\Gamma}}\rangle(t)  \right|  I_{sym}(\mathbf{s}(0))$.
In terms of regimes we thus expect: (i) $I_{sym} >0$; (ii) $I_{sym}=0$; (iii) $I_{sym}=0$.
  \end{itemize}

  Altogether, the estimations and calculations above lead to the following predictions for the presence of symmetry and structure as a function of time $t$ (regimes i-iii):
  \begin{itemize}
  \item[(i)] $0 \le t \le t_{metastable} = N/L:$\\
  Structure $I_{str} >0$ but no symmetry $I_{sym} > 0$.
  \item[(ii)] $ t_{metastable} = N/L \le t \le t_{equilibrium}=N$\\ 
  Structure $I_{str} >0$ and symmetry $I_{sym} = 0$.
  \item[(iii)] $  t_{equilibrium}=N < t;$\\ 
  Symmetry $I_{sym} = 0$ but no structure $I_{str} = 0$.
  \end{itemize}
In Fig.~\ref{fig2} we confirm these predictions in a numerical simulation.
\paragraph{The metastable regime.} The crucial feature of the TE dynamics discussed above is that in regime (ii) both non-trivial structure and symmetry co-exists in the generated sequences. The time (measured in number of iterations) for which this regime is valid is orders of magnitude larger than that of the first regime, as  the ratio $t_{equilibrium}/t_{metastable}=L$  corresponds to the average size of transposable elements (for example $L \simeq 10^2 $ in Homo Sapiens \cite{CBCHO07}). We thus denote such long-lived regime as \emph{metastable} and we expect it to be generically observed, even though it does not correspond to the stable equilibrium of our model.

\begin{figure}[t!]
\begin{center}
\includegraphics[width=\columnwidth]{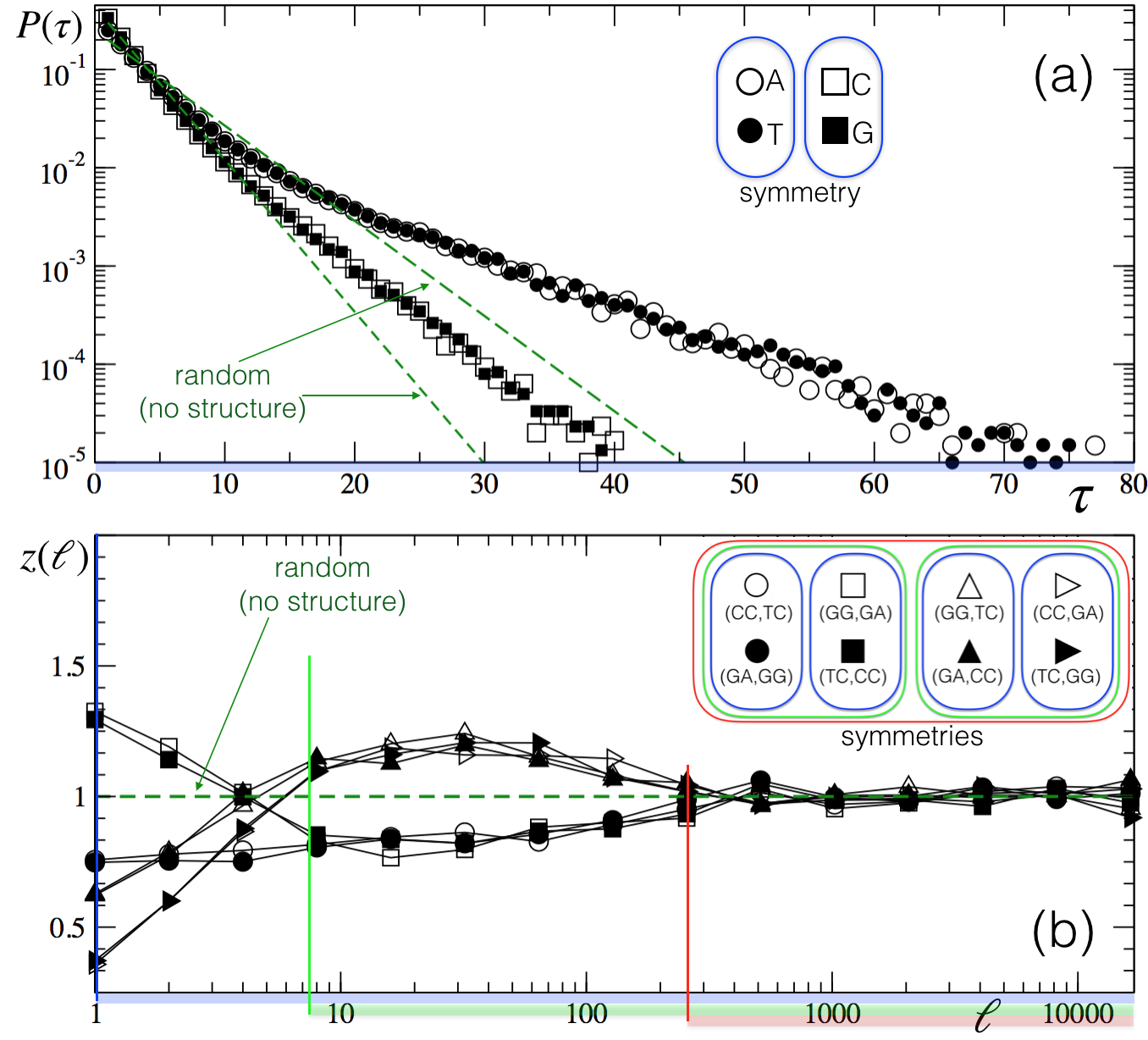}
\caption{{\bf{Symmetry and structure in the metastable regime.}}
Same observables as in Fig~\ref{fig.1} are computed for a sequence in the metastable regime of our dynamics.  Data show that this regime is characterized by  a similar  co-occurrence of symmetry and structure as in extant genomes.
Results in panel (a) are for a sequence of length $N=5 \times 10^6$ initialized as in Fig.~\ref{fig2} and evolved using our model with TE sizes $\ell $ all equals to $L=5000$ until $t=2048   \gtrapprox t_{metastable}=1000$. Results in panel (b) are for a sequence of length $N=10^5$ initialized as in the artificial sequence reported in Ref.~\cite{CDEA18} and evolved using our dynamic model with fixed $L=500$ until  $t=256 \gtrapprox t_{metastable}=200$. 
 The more generic initial sequence in panel (b) (i.e., Markov chain instead of fully random) allow us to distinguish between the different types of scale-dependent symmetries generated by the dynamics.
}
\label{fig3}
\end{center}
\end{figure}

\begin{figure}[t!]
\begin{center}
    \includegraphics[width=\columnwidth]{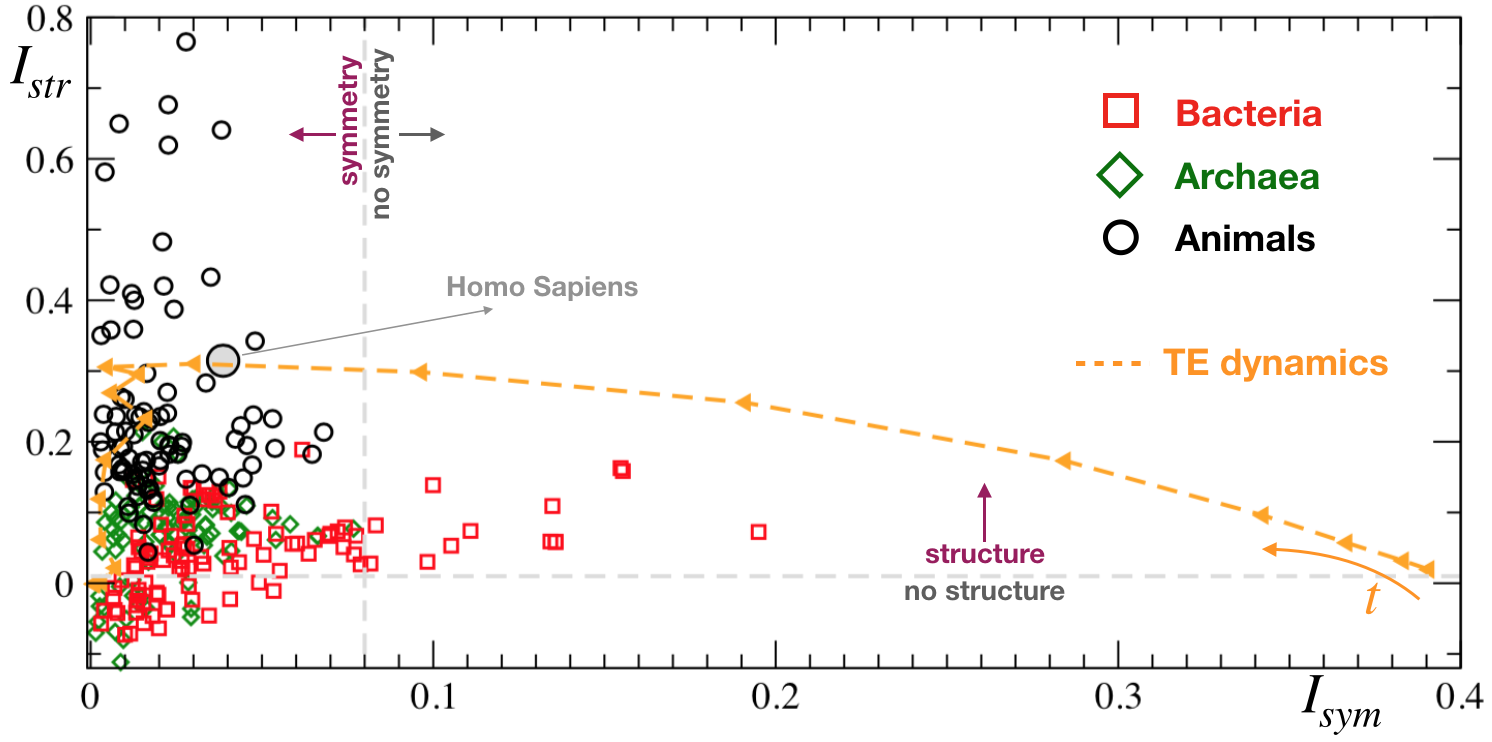}
\caption{{\bf 
Structure and symmetry in different organisms.} Values of $I_{sym}$ vs. $I_{str}$ for different genomes belonging to the families Archaea, Bacteria, Animals\cite{SM}.  Superimposed are the values of the sequences evolved via  our model (starting in $(I_{sym},I_{str})=(0.39,0)$ and evolving to $(0,0)$; parameters as in Fig.~\ref{fig2}a).} 
\label{fig4}
\end{center}
\end{figure}

The DNA sequences  in the metastable regime are characterized by a symmetric domain-like structure. Domain models have been already introduced in literature to reproduce the complex  structure generically observed in extant DNAs \cite{BGHPSS93,N92, KB93,  PBHSSG93, BGRO96,BOFZSCMR85,ARLR02,CBCHO07}. In particular if the distribution of domain sizes  has a fat tail, this will lead to a long-range correlated sequence \cite{BGHPSS93}, signalled by a slow decay of $P(\tau)$. The novelty of our approach is twofold: firstly,  the domain-like structure in the metastable regime
is an emergent property of the TE dynamics (it is not imposed a priori); secondly, such complex structure is intertwined with symmetry, that itself is an output of the dynamics.  In particular, we have shown that sequences in the metastable regime are not only Chargaff symmetric ($I_{sym}=0$),  they reproduce the  hierarchical relation between  symmetry and structure that is a distinctive feature of extant genomes (see  Fig.~\ref{fig3} ).
\paragraph{Different organisms.}
In Fig.\ref{fig4} we  report $I_{sym}$ and $I_{str}$ computed for genomes of different families, together with the values obtained from our dynamics. It shows that symmetry and structure coexist in most cases. The sequences from Animals shows enhanced structure while the cases of Archaea and Bacteria shows a moderate signatures of structure, in agreement with the temporal behaviour of our model (i.e., associating $t$ with the age of the genomes). 
 Note that symmetry  and structure  properties are both statistical observations we made on the full DNA sequence. Any evolutionary constraint that pertains a small percentage of an organism genome does not affect these statistical observation in a sensible way.  As an example, the protein-coding regions of Homo-Sapiens  account for $~1.5\%$ of the full sequence. On the other hand, care should be taken when dealing with many different organisms: extensions of the model incorporating additional aspects of DNA evolution will be required for a  quantitative comparison with the empirical data.
\paragraph{Conclusion.}  We have shown how a model that captures the action of transposable elements (TEs) is able to reproduce the intricate relation between symmetry and structure present in DNA sequences. We find that symmetry and structure change differently at different time scales (i.e., for different number of actions of TEs). For a large (pre-asymptotic) time interval, the sequences obtained in our model show the same non-trivial structures and an hierarchy of symmetries (including Chargaff) as in actual DNA sequences  (confront panels (b) of Fig.\ref{fig.1} and Fig.\ref{fig3}). Our mathematical model is extremely simplified and includes  the essential elements to explain the onset of symmetry and structure. In particular, it mimics only a simple action of TEs (reverse-complement), ignoring the fact that TEs are classified in different families, have different properties, and act according to different mechanisms \cite{JS88,munoz,JBK}.  
We expect that incorporating more details of the TE dynamics in our model will refine our understanding of their role in shaping statistical properties of DNA sequences, in particular in an evolutionary viewpoint that would lead to refinements in the data-model comparison presented in Fig.~\ref{fig4}.


\newpage
\begin{thebibliography}{99}
\bibitem{MC50} McClintock B,  The origin and behavior of mutable loci in maize, {\it  Proc. Natl. Acad. Sci. USA}  {\bf 36 }(6): 344-55 (1950).
\bibitem {FP07} Feschotte C, Pritham E.J., DNA Transposons and the Evolution of Eukaryotic Genomes, { \it Annu. Rev. Genet} {\bf 41}: 331-368 (2007).
\bibitem {K81}  Kleckner N, Transposable elements in prokaryotes, {\it Annu. Rev. Gen.} {\bf 15}: 341-404 (1981) 
\bibitem{BBGGetal18} Bourque G, Burns KH, Gehring M, Gorbunova V, Seluanov A, Hammell M, Imbeault M, Izsvák Z, Levin HL, Macfarlan TS, 
Mager DL, Feschotte C,  Ten things you should know about transposable elements,  {\it Genome Biology}  {\bf 19} (1): 199 (2018).
\bibitem{F12}  Fedoroff N.V.,  Transposable Elements, Epigenetics and Genome Evolution, {\it Science} {\bf 338}, 758-767 (2012).
\bibitem{HGBSH} Holste D, Grosse I,  Beirer S,  Schieg P, Herzel H, Repeats and correlations in human DNA sequences {\it Phys. Rev. E} {\bf 67}(6): 061913, (2003).
\bibitem{SRMA16} Sheinman, M, Ramisch, A,  Massip, F,  Arndt PF, Evolutionary dynamics of selfish DNA explains the abundance distribution of genomic subsequences,  {\it Scientific Reports} 6 (2016).
\bibitem{MA13} Massip F,  Arndt PF,  Neutral Evolution of Duplicated DNA: An Evolutionary Stick-Breaking Process Causes Scale-Invariant Behavior, {\it Physical Review Letters} {\bf 110} (14): 148101  (2013).
\bibitem{MAL05} Messer PW,  Arndt PF,  L\"{a}ssig M, Solvable Sequence Evolution Models and Genomic Correlations, {\it
Physical Review Letters} {\bf 94} 138103  (2005)
\bibitem{HJ96} Attard G, Hurworth A, Jack J, Language-like features in DNA: transposable element footprints in the genome {\it EPL (Europhysics Letters)} {\bf 36}, 391 (1996).
\bibitem{BGHPSS93} Buldyrev SV , Goldberger AL,  Havlin S, Peng CK, Simons M, and Stanley HE, Generalized Levy Walk Model for DNA Nucleotide Sequences, {\it Phys. Rev. E} {\bf 47}, 4514-4523 (1993),
\bibitem{B06}  Albrecht-Buehler  G,  Asymptotically increasing compliance of genomes with Chargaff's second parity rules through inversions and inverted transpositions, {\it Proc. Natl. Acad. Sci. USA}  {\bf 103}, 17828-17833 (2006).
\bibitem{RKC68}  Rudner R,  Karkas JD,  Chargaff E, Separation of B. subtilis DNA into complementary strands I. Biological properties, II. Template functions and composition as determined, III Direct analysis. {\it Proc. Natl. Acad. Sci. USA} {\bf 60}, 630-635; 915-922 (1968). 
\bibitem{R91} Rogerson AC, There appear to be conserved constraints on the distribution of nucleotide sequences in cellular genomes. {\it  J. Mol. Evol} {\bf 32}, 24-30 (1991). 
\bibitem{MB06} Mitchell D, Bridge R,  A test of Chargaff's  second rule  {\it Biochem. Biophys. Res. Commun.} {\bf 340}, 90-94  (2006).
\bibitem{NA06} Nikolaou C,  Almirantis  Y,  Deviations from Chargaff's second parity rule in organellar DNA Insights into the evolution of organellar genomes, {\it Gene} {\bf 381}, 34-41 (2006).
\bibitem{QC01} Qi D,  Cuticchia AJ,   Compositional symmetries in complete genomes, {\it Bioinformatics} {\bf 17}, 557-559 (2001).
\bibitem{FTW92} Fickett JW ,  Torney DC,  Wolf DR  Base compositional structure of genomes. {\it Genomics} {\bf 13}, 1056-1064 (1992).
\bibitem{P93}  Prabhu VV, Symmetry observations in long nucleotide sequences, {\it Nucleic Acids Res.}  {\bf 21}, 2797-2800 (1993).
\bibitem{BF99}  Bell SJ,   Forsdyke DR,  Accounting units in DNA. {\it J. Theor. Biol.}  {\bf 197}, 51-61 (1999).
\bibitem{PHB02}  Baisn\'ee PF,  Hampson S,  Baldi P, Why are complementary DNA strands symmetric?, {\it Bioinformatics} {\bf 18}, 1021-1033 (2002).
\bibitem{KFCHZZL09}   Kong S-G,  Fan W-L, Chen H-D,  Hsu Z-T, Zhou N,   Zheng Bo, Lee H-C,  Inverse Symmetry in Complete Genomes and Whole-Genome Inverse Duplication, {\it PLOS one} {\bf 4}, e7553 (2009).
\bibitem{ABGRPF13}  Afreixo V1, Bastos CA, Garcia SP, Rodrigues JM, Pinho AJ, Ferreira PJ,  The breakdown of the word symmetry in the human genome, {\it J. Theor. Biol.}  {\bf 335}, 153-1599 (2013).
\bibitem{BF99b}   Bell SJ,  Forsdyke DR, Deviations from Chargaff's Second Parity Rule Correlate with Direction of Transcription, {\it J. Theor. Biol.}  {\bf 197},  63-76 (1999).
\bibitem {LL99}  Lobry JR,  Lobry C,  Evolution of DNA base composition under no-strand-bias condition when the substitution rates are not constant, {\it  Mol. Biol. Evol.}  {\bf 16}, 719-723 (1999).
\bibitem{ZH10}  Zhang SH,  Huang YZ,  Limited contribution of stem-loop potential to symmetry of single-stranded genomic DNA, {\it Bioinformatics} {\bf 26}, 478-485 (2010).
\bibitem{HMO12}Hart A, Mart\'inez S, Olmos FA, Gibbs Approach to Chargaff’s Second Parity Rule, {\it Journal of Statistical Physics} {\bf 146}, 408-422 (2012).
\bibitem{CBHDK19} Coons LA, Burkholder AB, Hewitt SC, McDonnell DP, Korach KS,  Decoding the Inversion Symmetry Underlying Transcription Factor DNA-Binding Specificity and Functionality in the Genome, {\it  iScience}  {\bf 15}, 552-591 (2019)
\bibitem{FTPM19} Fariselli P, Taccioli C, Pagani L, Maritan A, DNA sequence symmetries from randomness: the origin of the Chargaff’s second parity rule, {\it Briefings in Bioinformatics}, bbaa041, (2020).
\bibitem{PBGHSSS92} Peng CK ,  Buldyrev SV,Goldberger  AL ,  Havlin S,  Sciortino F, Simons M  and Stanley HE,  Long-range correlation in nucleotide sequences, {\it Nature} {\bf 356}, 168-170 (1992).
\bibitem{LK92}  Li W, Kaneko K,  Long-Range Correlation and Partial $1/f^{\alpha}$ Spectrum in a Noncoding DNA Sequence, {\it EPL} {\bf 17}, 655-660 (1992).
\bibitem{V92} Voss R,  Evolution of Long-Range Fractal Correlations and $1/f$ Noise in DNA Base Sequences, {\it Phys. Rev. Lett.} {\bf 68}, 3805-3808  (1992).
\bibitem{A92}  Amato I,  DNA shows unexplained patterns writ large, {\it Science} {\bf 257}, 747(1992).
\bibitem{P92} Yam P,  Noisy nucleotides: DNA sequences show fractal correlations, {\it Sci. Am.} {\bf 267}, 23-24,27 (1992).
\bibitem{LMK94}  Li W, Marr TG,  Kaneko K,  Understanding long-range correlations in DNA sequences, {\it Physica D} {\bf 75}, 392-416 (1994).
\bibitem{ATVAMA01} Audit B, Thermes C, Vaillant C, d'Aubenton-Carafa J,  Muzy JF and Arneodo A, Long-Range Correlations in Genomic DNA: A Signature of the Nucleosomal Structure, {\it Phys. Rev. Lett.} {\bf 86}, 2471 (2001).
\bibitem{FS12} Frahm  KM, Shepelyansky DL, Poincar\'e recurrences of DNA sequences, {\it Phys. Rev. E} {\bf 85}, 016214 (2012).
\bibitem{CPTC15} Colliva, A,  Pellegrini R, Testori A, Caselle M,  Ising-model description of long-range correlations in DNA sequences, {\it Phys. Rev. E}  {\bf 91}: 052703 (2015).
\bibitem{CDEA18} Cristadoro G, Degli Esposti M,  Altmann EG,  { The common origin of symmetry and structure in genetic sequences}, {\it Scientific Reports} {\bf 8}(1), 15817 (2018). 
\bibitem{N92}  Nee S,  Uncorrelated DNA walks, {\it Nature} {\bf  357}, 450 (1992).
\bibitem{KB93} Karlin S, Brendel V,   Patchiness and correlations in DNA sequences, {\it Science} {\bf 259}, 677-680 (1993).
\bibitem{PBHSSG93} Peng CK, Buldyrev SV, Havlin S, Simons M, Stanley HE and Goldberger AL,  Mosaic organization of DNA nucleotides, {\it Phys. Rev. E}  {\bf 49}, 1685-1689 (1994).
\bibitem{BGRO96} Bernaola-Galv\'an P, Rom\'an-Rold\'an R,  Oliver JL,  Compositional segmentation and long-range fractal correlations in DNA sequences, {\it Phys. Rev. E}  {\bf 53}, 5181-5189  (1996).
\bibitem{BOFZSCMR85}  Bernardi G, Olofsson B, Filipski J, Zerial M, Salinas J, Cuny G, Meunier-Rotival M and Rodier F, The mosaic genome of warm-blooded vertebrates, {\it Science } {\bf 228}, 953-958 (1985).
\bibitem{ARLR02} Rajeev K, Azad J, Subba R, Wentian Li, and Ramakrishna R, Simplifying the mosaic description of DNA sequences, {\it Phys. Rev. E} {\bf 66}, 031913 (2002).
\bibitem{CBCHO07} Carpena P ,  Bernaola-Galv\'an P,  Coronado AV,  Hackenberg M,  Oliver JL, Identifying characteristic scales in the human genome, {\it Phys. Rev. E} {\bf 75},  032903 (2007).
\bibitem{JS88} Jurka J, Smith T, A fundamental division in the Alu family of repeated sequences, {\it Proc Natl Acad Sci U S A} {\bf 85}, 4775–8 (1988).
\bibitem{munoz} Muñoz-López M,  García-Pérez JL,  DNA transposons: nature and applications in genomics, {\it Current genomics} {\bf 11}(2), 115–128 (2010).
\bibitem{JBK} Jurka J, Bao W, Kojima KK, Families of transposable elements, population structure and the origin of species, {\it Biology direct} {\bf 6}, 44 (2011).
\bibitem{SM} All sequences were downloaded from the National Center for Biotechnology Information (https://www.ncbi.nlm.nih.gov/). The sequences were processed to remove all letters different from A, C, G, T. The first 100000 subsequence of each entry was used for data in Fig.4.
\end{thebibliography}
\end{document}